\documentclass[twocolumn,aps,pra,floatfix,showpacs,noshowkeys,epsfig,graphics,natbib]{revtex4}
\usepackage{graphicx}
\usepackage{amsmath}
\usepackage{amsfonts}
\usepackage{amssymb}

\newcommand{\newc}{\newcommand}
\newc{\beq}    {\begin{equation}}
\newc{\eeq}    {\end{equation}}

\newc{\beqa}    {\begin{eqnarray}}
\newc{\eeqa}    {\end{eqnarray}}
\newc{\bs}    {\section}
\newc{\no}    {\\ \nonumber}

\newc{\st}    {\stackrel}

\begin{document}
\title{ Quantum Informational Dark Energy: Dark energy from  forgetting}
\author{Jae-Weon Lee}\email{scikid@kias.re.kr}
\affiliation{School of Computational Sciences,
             Korea Institute for Advanced Study,
             207-43 Cheongnyangni 2-dong, Dongdaemun-gu, Seoul 130-012, Korea}

\author{Jungjai Lee}
\email{jjlee@daejin.ac.kr}
\affiliation{Department of Physics, Daejin University, Pocheon, Gyeonggi 487-711, Korea}

\author{Hyeong-Chan Kim}
\email{hckim@phya.yonsei.ac.kr}
\affiliation{Center for Quantum Spacetime, Sogang University,
Seoul 121-742, Republic of Korea.
}%
\date{\today}
\begin{abstract}
We suggest that dark energy has a quantum informational origin. Landauer's principle
associated with the erasure of quantum information at a cosmic  horizon
 implies  the non-zero vacuum energy having
effective negative pressure.
Assuming the holographic principle, the minimum free energy  condition, and
the Gibbons-Hawking temperature for the cosmic event horizon
 we obtain
 the holographic dark energy with the parameter $d\simeq 1$,
 which is consistent with the current observational data.
It is also shown that both the entanglement energy and the horizon energy
can be related to Landauer's principle.
\end{abstract}

\maketitle
\section{Introduction}
The cosmological constant problem is one of the most important unsolved puzzles in modern
physics~~\cite{CC}.
There are strong evidences from  Type Ia supernova (SN Ia) ~~\citep{riess-1998-116,perlmutter-1999-517},
cosmic microwave background radiation~\cite{wmap3} and
Sloan digital sky survey (SDSS) observations~\cite{SDSS1}
 that
the current universe is under an accelerating expansion, which can be
explained  by dark energy (a generalization of the cosmological constant) having  energy density $\rho_\Lambda$
and negative pressure
 $p_\Lambda$  satisfying
$w_\Lambda\equiv p_\Lambda/\rho_\Lambda<-1/3$.
Although, there are already various models relying on  materials such as
quintessence~\cite{PhysRevLett.80.1582,PhysRevD.37.3406},
$k$-essence~~\cite{PhysRevLett.85.4438}, phantom~~\cite{phantom,nojiri-2006-38}, Chaplygin gas~~\cite{Chaplygin},
and quintom~\cite{quintom}
among many~\cite{wei-2007,Gough:2007cj,Copeland:2006wr},
the identity of the dark energy is still mysterious.
There are also many attempts to derive dark energy
from the  quantum vacuum fluctuation (See ~\cite{Zeldovich,Gurzadyan:2000ku,Padmanabhan:2007xy,Mahajan:2006mw,Djorgovski:2006vn}).
A fundamental problem
in all these models is that there is no obvious way to cancel the $O(M_P^4)$ zero point energy from the quantum vacuum  fluctuation.
Here $M_P$ is the reduced Planck's mass.
On the contrary,  the holographic dark energy (HDE) model
has  an intrinsic advantage over other  models in that
it does not need fine-tuning of parameters or  an  $ad~hoc$ mechanism to
cancel the zero-point energy of the vacuum,  because quantum fields
have less degrees of freedom from the start in this model,
according to the holographic principle~\cite{holography}.
Furthermore, in this model, the cosmic coincidence problem
could  be also solved  if there was an inflation with the number of efolding $N\simeq 65$  \cite{li-2004-603,mycoin}.
 However, HDE model has its own difficulties~\cite{ageproblem,Instability,Li:2008zq} and the
  physical origin of HDE itself has yet to be
 adequately justified.
Recently, we suggested that there can be a
 relation between dark energy and entanglement (nonlocal quantum correlation)
 which are the two great puzzling entities in modern physics~\cite{myDE}.

In this paper, we suggest a new idea that dark energy is originated from the erasure of
the quantum information
at the cosmic  horizon.
This idea linking dark energy to quantum informational concepts is not so strange than it may appear,
 because we can treat every physical process in the universe as essentially a kind
of quantum information processing~\cite{Lloyd2000,nielsen,Plenio1999}.
We will show that Landauer's principle
applied to this cosmological information erasing implies the existence of non-zero vacuum energy having
effective negative pressure. Being one of main concepts  in quantum information science~\cite{landauer,PhysRevA.61.062314}, Landauer's principle states that
to erase one bit information of a system irreversibly at least $k ln 2$ entropy should be increased
and at least $k T ln 2$ free energy should be consumed,
where $k$ is the Boltzman's constant and $T$ is the temperature of a thermal bath
contacting with the system. (Henceforth, we set $k=1$.)
Since this erased information can be connected to the vacuum fluctuation, our model
provides a new way to obtain dark energy from the vacuum fluctuation.
Bennett~\cite{maxwell} solved the   Maxwell's demon problem using this principle,
which is also related to  reversible computation~\cite{nielsen} and quantum computation.
Landauer's principle have been also used to study
 thermodynamics of black holes by many authors~\cite{song-2000,braunstein-2007-98,Hosoya:2000yb,Kim:2007vx,MohseniSadjadi:2007zq}.
We will also show that the entanglement energy and the horizon energy
can be explained by Landauer's principle.


In Sec. II we show that the minimum free energy  condition and
the holographic principle give a rise to HDE with $d\simeq 1$ as observed.
In Sec. III, we review Landauer's principle and suggest that
this minimum free energy  condition is related to the quantum information erasing
at the cosmic horizon and dark energy comes from this principle.
 In Sec. IV, we consider more general situations by relaxing the restrictions for dark entropy( the entropy related to dark energy) $S_\Lambda$ and
 the horizon temperature $T$. We show that even in this
  case the quantum informational dark energy has a form of HDE generally.
Section V contains discussions.

\section{Holographic dark energy and Minimum Free energy condition}
In this section we show that
by
assuming the holographic principle, the minimum free energy condition and
Gibbons-Hawking temperature for the cosmic event horizon,
we can easily obtain
 the  HDE with the parameter $d\simeq 1$ as observed.
 In section II and III we choose the event horizon as an IR-cutoff.
  Before going further
 we need to shortly review the entanglement dark energy model~\cite{myDE}.
The vacuum entanglement energy ~\cite{entenergy}
associated with the entanglement
entropy  $S_{Ent}$ ~\cite{Srednicki} between inside and outside parts of the cosmic event horizon
could give dark energy density
in the form of HDE
~\cite{li-2004-603}
 \beq
 \label{holodark2}
\rho_\Lambda=\frac{3 d^2 M_P^2}{ R_h^2 },
\eeq
where $R_h$ is the radius of the future event horizon
and $d$ is a parameter.
Obtaining the exact value of $d$ is important, because  it
determines the properties of HDE and
 its approximate value was derived recently only in~\cite{myDE}.
More precisely, we suggested that (entanglement) dark energy satisfies
\beq
\label{eenergy}
dE_{\Lambda}= T dS_{\Lambda},
\eeq
where the temperature $T$  is
 the Gibbons-Hawking temperature of the event horizon
~\cite{PhysRevD.15.2738,izquierdo-2006-633,1475-7516-2006-09-011} and
the dark entropy
 $S_\Lambda$ is given by the entanglement entropy of the vacuum fluctuation
 defined as
 \beq
 \label{Sent}
  S_{Ent}(\rho_A)\equiv-Tr(\rho_A log \rho_A)= h (\rho_A),
  \eeq
  where $h$ is the von Neumann entropy.
   This energy can be also interpreted as the `horizon energy'~\cite{BoussoDesitter} representing
 the vacuum energy inside the horizon according to the holographic principle.
 It is natural  to divide the universe $(AB)$ into
 the inside ($A$) and the outside ($B$) of the event horizon  to calculate the entanglement
 entropy, because the event horizon represents
the global causal structure. (See Fig. 2.)
 The reduced density matrix
$\rho_A\equiv Tr_B \rho_{AB}$ represents an effective subsystem $A$
 of the whole system $AB$ described by a density matrix $\rho_{AB}$~\cite{nielsen,Hubeny:2007xt}.
This entanglement
 dark energy can
 be also related to the cosmic Hawking radiation (See the comments in ~\cite{newscientist,sciencenews}).

Using  $M_P$ as a UV cut-off and the spin degrees of  freedom ($N_{dof}=118)$
in the standard model,
we obtained the
parameter (Eq. (12) in Ref. \cite{myDE})
 \beq
\label{d1}
 d=\frac{\sqrt{0.3 N_{dof}}}{2\pi}\simeq 0.95,
 \eeq
 which is well consistent with  current observational
 data $d=0.91^{+0.26}_{-0.18}$~\cite{zhang-2007}.
Although the chosen values for these input parameters seems to be reasonable, it is desirable to get a more
 plausible and general model having
less free parameters, which we pursue in this paper.

To remove the ambiguity of input parameters, in this
section, we assume  that the quantum informational
dark entropy $S_\Lambda$ is given by the holographic
principle ~\cite{PhysRevD.15.2738}, i.e.,  $S_\Lambda=S_{hol}\equiv A m_P^2/4=\pi R_h^2 m_P^2$, where $A\equiv 4\pi R_h^2$ is
the surface area of the cosmic event horizon and $m_P=\sqrt{8 \pi} M_P$.
For the temperature $T$  we use
the Gibbons-Hawking temperature~\cite{PhysRevD.15.2738} $T=1/2 \pi R_h$.
 This is a plausible choice because
our universe is under an accelerating expansion and going to a dark energy dominated universe,
which can be a quasi-de Sitter universe.(This kind of space-time attracts much interest recently in relation with dS/CFT correspondence ~\cite{Medved:2001ne}.)
Note that although this temperature is very low, due to the huge entropy proportional
 to the horizon area, dark energy $E_\Lambda\sim T S_\Lambda$ could be comparable to the
  observed value. By inserting $S_\Lambda$ and $T$ into Eq. (\ref{eenergy})
 and integrating it one can obtain
 \beq
 \label{int}
 E_\Lambda=\int dE_\Lambda=\int T dS_\Lambda= 8\pi R_h M_P^2
 \eeq
 and HDE density $\rho_\Lambda=E_\Lambda/(4\pi R_h^3/3)=6 M_P^2/R_h^2$ with
$d=\sqrt{2}$, which, however, somewhat deviates from the observed value
$d=0.91^{+0.26}_{-0.18}$.
This problem can be alleviated by considering
 the minimum free energy condition
$dF=d(E_\Lambda-T S_\Lambda)=0$, or,
\beq
\label{dE2}
dE_\Lambda=d(T S_\Lambda)=T dS_\Lambda + S_\Lambda dT ,
\eeq
instead of Eq. (\ref{eenergy}).
The second term may represent the contribution from the change of the temperature
in addition to the typical thermal energy term $T dS_\Lambda $.
(In section III we will consider another interpretation of this minimum free energy condition.)
In this case
\beq
\label{dE4}
dE_\Lambda=d\left(\frac{\pi R_h^2m^2_P}{2\pi R_h}\right)=\frac{m^2_P dR_h }{2},
\eeq
from which  one can obtain $d=1$ by repeating the integration in Eq. (\ref{int}).
For $d=1$ the equations (\ref{omega}) and (\ref{omega3}) below  give
the  equation of state $w^0_\Lambda=-0.903$ and its derivative $w_1=0.208$
at the present. (See  Eq. (\ref{w1}) for the exact definition of $w_1$.)
These results are comparable to  the recent observational data;
$w^0_\Lambda = -1.03 \pm 0.15$ and  $w_1 = 0.405^{+0.562}_{-0.587}$ ~\cite{xia-2007,Zhao:2006qg}.
(For $d=\sqrt{2}$,  $w^0_\Lambda=-0.736$ and $w_1=0.12$.)
This is our first main result, which suggests that  the correct equation for  the dark energy  could be Eq. (\ref{dE2}) rather than Eq. (\ref{eenergy}).
This seems to be reasonable, because we need to consider the effect of the variation of $T$ on $E_\Lambda$.

Since our universe is not exactly equal to the de Sitter universe, $T$ can be
slightly different from that of the  de Sitter universe. Thus, it is also expected that
$d$ is approximately 1  as observed.
Recall that we get  this result without introducing ambiguous parameters such as the UV-cutoff or $N_{dof}$
in our calculation. With the simple and reasonable assumptions, HDE with $d\simeq 1$
can be easily derived.

It was shown in ~\cite{li-2004-603} that if the event horizon of  our universe is that of a black hole,
using the black hole
Hawking temperature $T=1/4 \pi R_h$ and $dE_\Lambda=T dS_\Lambda$
one can obtain $E_\Lambda=4\pi R_h M_P^2$
and $\rho_\Lambda=3 M_P^2/R_h^2$, that is,  $d=1$.
However, it is obvious that our universe is not exactly like a black-hole.

It is tempting to interpret the minimum free energy   condition in Eq. (\ref{dE2}) as the condition for the vacuum in
a canonical ensemble, however, it is unclear whether we can treat the horizon as a thermal system in a canonical ensemble.
In the next section we consider an alternative possibility that this condition
comes from Landauer's principle and dark energy has a quantum informational origin.

\section{Landauer's Principle and dark energy}

First of all,
it is important to make a clear distinction between two terminologies, the information `loss' in black hole physics
and the information `erasing' in quantum information science.
By information `loss' we mean, for example, a non-unitary transformation from a pure state to
a mixed thermal state at a black hole.
On the other hand the information `erasing' in quantum information science is a transformation from unknown states
to a specific  state by compulsion, a process usually $reducing$ the entropy of the state ~\cite{PhysRevA.61.062314}.
It is also different from `measurements' where the unknown states randomly collapse to one of measurement basis states.
For example, consider  an one-bit  memory which consists of cylinder and an atom in it as shown in Fig. 1.
A thermal bath
 with the temperature $T$  keeps in contact with the cylinder.
Let us denote `0' (`1') by locating the atom in the left (right) partition.
To erase this memory we use the piston to push the atom into the left partition
regardless of its initial position by investing a work $W_{sys}$.
Then, the initial information of the atom is erased irreversibly, which is similar
to what happens during a formatting of a computer hard disk.
According to  Landauer's principle the entropy of the cylinder $decreases$ by $ ln 2$ at the cost of
 at least $T ln 2$ free energy
consumption, which means $W_{sys}\ge T ln 2$.
(See \cite{plenio-2001-42,20000408,2005qai..book.....S} for details.)
This energy is eventually  converted to thermal energy of the bath increasing the total entropy of the $whole$ system  (the  bath+
the cylinder) and  saving the second law of thermodynamics.
Note that, for general systems, $W_{sys}$ needs not to be a mechanical work.
Any free energy that can be used to erase the information and to reduce free energy of a system
can play a similar role.

\begin{figure}[thbp]
\includegraphics[width=0.32\textwidth]{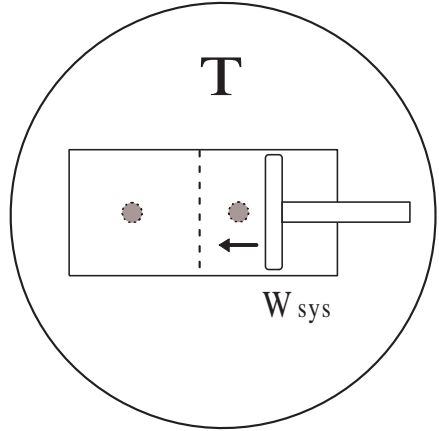}
\caption{According to Landauer's principle, to erase one bit information (the position of an atom, the gray points) in the cylinder at least $Tln2$
work $W_{sys}$ should be invested. Note that the local entropy within the cylinder $decreases$ while  the entropy
of whole system (cylinder+bath)
should not decrease.
 \label{ccorrelation} }
\end{figure}
More generally, according to the principle to erase the information of a system (for example, the cylinder in Fig. 1)
represented by $dS_{sys}<0$, the work   $W_{sys}\ge  T |dS_{sys}|$  should be invested, if $T$ is fixed.
 This eventually  increases the entropy of the bath  by $dS_{bath}$
 and the thermal energy of the bath   at least by $T|dS_{sys}|$  such that the total entropy
  increases by $dS_{tot}\equiv dS_{bath}+ dS_{sys}\ge 0$.
  Alternatively, Landauer's principle can be  restated as the claim
 that the net free energy gain of the whole system during the erasure
can not be larger than  the  work invested, i.e.,
  \beqa
  \label{dF}
&dF_{tot}&\equiv dF_{bath}+dF_{sys} \no
&=&d(E_{bath}-T S_{bath})+d(E_{sys}-T S_{sys})\le W_{sys}.
\eeqa
Otherwise, one can use the increased net free energy $dF_{tot}$ to run a
 perpetual motion machine of the second kind,
 because
  the free energy is a measure of the amount of work  extractable  from a system.
  This   clearly violates
 the second law of thermodynamics.
 One can say that the erasure is done most efficiently when the above  equality holds.
 In this case
  the work $W_{sys}$ is used solely to erase the information and the net increase of the total entropy is zero.
 For this optimal case and $T$ fixed,
 \beq
 dS_{bath}=|dS_{sys}|,~dE_{sys}=0,~
  W_{sys}=T|dS_{sys}|=dF_{sys},
 \eeq
 and the condition in Eq. (\ref{dF}) reduces to $dF_{bath}=0$, the minimum free energy condition,
 which leads to $dE_{bath}=TdS_{bath}$.
 This suggests that  the  the horizon energy ~\cite{BoussoDesitter}
 or the `first-law' often given in this form
 has a quantum informational origin.
 In Ref. \cite{Kim:2007vx}, by applying a similar condition to the black hole horizon, we derived the first law of black hole physics and the discrete black hole mass.
 For more general cases where
 $T$ can vary,
we can expect that  $W_{sys}=dF_{sys}$ and
 the minimum free energy condition ($dF_{bath}=0$) still
holds for an optimal erasing process, where information is erased most efficiently.

 In this paper, we suggest that the cosmic horizon plays the role
 of the bath and is one of the most efficient
  information eraser saturating the bound
in Eq. (\ref{dF}) and
  dark energy is this marginal horizon (vacuum) energy $E_\Lambda$
 originated from the quantum information erasing
at the cosmic  horizon with the Gibbons-Hawking temperature $T$.
 That is, $dF_{bath}=dF_\Lambda=0$.
 Thus, the change of dark energy $E_\Lambda$ is related to the increase of
 dark entropy by $dS_\Lambda$
at the horizon as
$dE_\Lambda=d(T S_\Lambda)$.

Let us discuss in detail how
one can apply  Landauer's principle to the cosmic horizon. (See Fig. 2.)
Here, the horizon plays a role of both the piston
and the bath in Fig. 1 at the same time.
For an observer $O_A$
the larger his/her cosmic horizon expands, the more information about
outer region  is erased at the horizon, because $O_A$
sees an expanding spherical horizon eating up the
outer space as it expands.
What exactly is the information erased here?
Let us consider the quantum field $\phi$ with a Lagrangian $L(\phi)$ in the universe.
Following Ref. \cite{Srednicki}
we can discretize the space and think the field
as a collection of linked
 quantum oscillators $\phi_i$ located on the lattice with
  a Planck length  ($l_p=1/m_P=\sqrt{G}$) spacing.
 Then, the vacuum is the ground state of the oscillators.
As the horizon expands, the inside region ($A$) is extended  by engulfing the outer region (the gray region in the figure).
Since the cosmological variation of the gravitational constant $G$ seems to be negligible ($|d G/dt/G|\st{<}{\sim}10^{-12}~yr^{-1}$) ~\cite{PhysRevLett.77.1432},
 we expect the enlargement of the region $A$ is not by
 the extension of the lattice spacing $l_P$ but by
 the appearance of new lattice sites and oscillators on them in the ground state
  at the horizon.
 The information erased is that of the new oscillators appeared
  at the region having been once outside the horizon (the gray region).
For the observer $O_A$, regardless of their initial quantum states,
the oscillators at the erased region are
 forced to go to a specific state (the ground state) of the region $A$ by compulsion
as the horizon expands. During this process the local entropy of the gray region $decreases$, while that of the
horizon increases.
(Note that this does not mean that normal particles outside the horizon can enter the region $A$.)
This is a kind of information erasure.
Since we are interested in the energy of the vacuum, for simplicity, we ignore the erasure
of the information of matter at the horizon.

For a toy example, imagine the black dot in the figure as a `vacuum' qubit in the gray region.
For the observer this vacuum qubit is initially  out of the horizon and can be described as the maximally
mixed state, $\rho=I/2$ having $ln2$ entropy. As the horizon expands, the region is reachable and
the qubit becomes
a part of the visible vacuum of the observer, which has zero entropy.
Thus, $\Delta S_{sys}=-ln2$. This `resetting' requires increase of the entropy of the thermal bath (horizon)
by $ln2$ .

\begin{figure}[hbtp]
\includegraphics[width=0.32\textwidth]{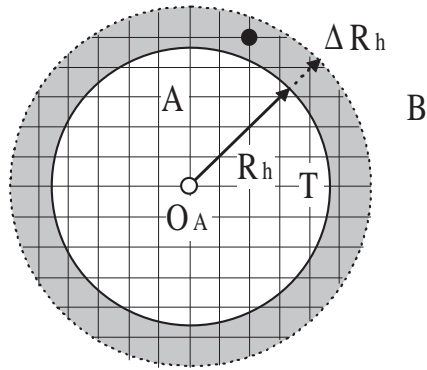}
\caption{
Expansion of the cosmic event horizon with a radius $R_h$ and
the Hawking temperature $T$
induces the information erasing of the system (the gray region) with $dS_{sys}$ at the horizon
for an inside observer $O_A$.
Regardless of its initial quantum state, the quantum field (the black dot) at the formerly unreachable region
is becoming a part of the reachable universe for the observer and
 forced to be the ground state
as the horizon expands.  Due to  Landauer's principle,
this information erasing consumes the free energy at least $|d(TS_{sys})|$, which
 can turn into dark energy finally.
 }
\end{figure}
Consider a more realistic situation, where
the horizon changes from $R_h$ to $R_h+\Delta R_h$
during the time interval $[t,t+\Delta t]$.
The density matrix of the erased system which is between
 $R_h$ and $R_h+\Delta R_h$
 changes from $\rho _{sys} (t)$
to $\rho _{sys} (t + \Delta t)$, and the total system changes
from $\rho _{AB} (t )$ to $\rho _{AB} (t+ \Delta t)$, while
 the dark entropy of the horizon $S_\Lambda$ changes  as
\beq
\Delta S_\Lambda  = S_\Lambda (\rho _{AB} (t + \Delta t)) - S_\Lambda(\rho _{AB} (t)).
\eeq
The horizon at $t$ plays a role of the heat bath for this time interval.
 For the optimal erasing,
  it should satisfy the minimum free energy condition;
$\Delta F_\Lambda=0$.
Thus, the increase of the horizon energy is $\Delta  E_\Lambda=\Delta (T S_\Lambda)$.
If this energy $E_\Lambda$ increases as the universe expands, it can play a role of dark energy
(See Eq. (\ref{p})).
This is a cosmological embodiment of  the famous slogan in quantum information science
``information is physical!".

Let us consider an infinitesimal erasing process.
In general, it is not easy to derive
$d S_{sys}$ directly. Fortunately, what we need to obtain $\rho_\Lambda$ is
 $dS_\Lambda$.
We already know two methods to calculate $dS_\Lambda$.
First, if this $S_\Lambda$ is the entanglement entropy $S_{Ent}(\rho_A)$~\cite{Srednicki}
in Eq. (\ref{Sent}), the model reduces to the entanglement
dark energy model in Ref. \cite{myDE} (with $S_\Lambda dT$ ignored), where
$dS_\Lambda (\rho_{AB}) =d h(\rho _{A})=d(-Tr(\rho_A ln \rho_A))$.
The information erased here is the entanglement information  of the vacuum fluctuation and calculated in Ref. \cite{Srednicki,myDE}.
In this case the dark entropy is from the entanglement of the vacuum fluctuation (virtual particles)
created near the horizon.
This explains the physical origin of the entanglement energy~\cite{entenergy}, which has many different definitions.
(Peacock had conjectured there is energy in entangled systems related to  Landauer's principle~\cite{peacock}.)
If we use Eq. (\ref{dE2})  instead of  Eq. (\ref{eenergy}) for the entanglement
 dark energy in Ref.~\cite{myDE},
 with the same UV-cutoff, one can obtain
 \beq
 d=\frac{\sqrt{0.3 N_{dof}}}{2\sqrt{2}\pi}\simeq 0.67
 \eeq
 and  $w_\Lambda^0\simeq -1.18$ for the  standard model
and  ($d\simeq 0.96 $, $w_\Lambda^0\simeq -0.925$) for the minimal supersymmetric standard model ($N_{dof}=244$).
That means the definition of dark energy in  Eq. (\ref{dE2}) seems to favor supersymmetry
when applied to the entanglement dark energy. (However, there are also arguments that the observations
favor $w_\Lambda^0$ smaller than $-1$~\cite{quintom}.)

 Second,
if we think $S_\Lambda$ as the entropy given by the holographic principle $S_{hol}=\pi R_h^2 m_P^2$, then we can easily obtain HDE with $d= 1$ using  Eq. (\ref{dE4}) as shown in the section II.
The information erased in this case can be interpreted as  the all degrees of freedom in the vacuum
in the gray region. This interpretation could explain the physical origin of the horizon energy
of the universe with the horizon.

Note that for both cases we get  $d$ values and dark energy density
explicitly. It is an important open question whether $S_{Ent}$ is exactly equal to $S_{hol}$ or not~\cite{1126-6708-2006-01-098}.

Contrary to the black hole case, the location of event horizons in the Friedmann universe
 depends on a choice of observer~\cite{Bousso:2002fq}.
  However, due to the cosmological principle
every point in the universe has the same amount of dark energy at the same comoving time.

\section{Dark energy for more general horizons}

In this section, we even further reduce the assumptions about the dark entropy, the horizon and the horizon temperature.
From information science viewpoint, the event horizon is the most natural candidate
for our purpose,
because it plays a role of a natural information barrier.
However, we need to check whether other horizons, such as Hubble horizon, particle horizon and
apparent horizon can replace the event horizon or not.
We adopt the approach of Hsu and Li~\cite{Hsu,li-2004-603} in this paper.
Since the Hubble horizon and the apparent horizon can be
related to the particle horizon asymptotically~\cite{Akbar:2006kj},
we can concentrate only on the future event horizon and the particle horizon.
Let us denote the radius of these horizons as $r$.

The particle horizon is defined as
\beq
\label{Rp}
R_p\equiv R(t)\int_0^t \frac{d t'}{R(t')}= R(t)\int_0^R \frac{d R(t')}{H(t') R(t')^2},
\eeq
and the future event horizon is
\beq
\label{Rh}
R_h\equiv R(t)\int_t^\infty \frac{d t'}{R(t')}= R(t)\int_R^\infty \frac{d R(t')}{H(t') R(t')^2},
\eeq
where $R(t)$ is the scale factor of the universe and $H$ is the Hubble parameter as usual.
Here we consider the flat ($k=0$) Friedmann universe which is favored by observations and inflationary theory
and is described by the metric
\beq
ds^2=-dt^2+R^2(t)d\Omega^2.
\eeq

We investigate the behavior of dark energy during the dark energy dominated era only.
In cosmology,  physical quantities are often expressed as  polynomial functions of some length scale.
For example, for the flat Friedmann universe energy density, scale factors, and comoving time can be
written as
power-law functions of the Hubble horizon size.
Let us assume, in this paper, that the dark entropy and the temperature are power-law functions
of some horizon radius ($r$), that is,  $S_\Lambda\propto r^\sigma$ and $T\propto r^\tau$.
For the entanglement dark energy $\sigma=2$ and $\tau=-1$. If this temperature is the Hawking temperature
our observable boundary of the universe is something like a black hole horizon.
By integrating $dE_\Lambda=d(T S_\Lambda)$ one can  obtain
\beq
\label{ELambda}
E_\Lambda=\epsilon M_P^{n+1} r^n,
\eeq
 where $n\equiv \tau+\sigma$ and $\epsilon$ is a dimensionless proportional constant which depends on the exact
definitions of $T$ and $S_\Lambda$.
Using $dE_\Lambda=T dS_\Lambda$ will give the same results up to a proportional constant.
(We will see later that this constant $\epsilon$ determines properties of dark energy.)
Thus, the dark energy density is given by
\beq
\label{rho}
\rho_\Lambda=\frac{3 E_\Lambda}{4 \pi r^3}=\frac{3\epsilon M_P^{n+1} r^{n-3}}{4 \pi}=3 M_P^2 H^2,
\eeq
where the last equality comes from the Friedmann equation.
From this we obtain
\beq
\label{r}
r= \eta H^{\frac{2}{n-3}},
\eeq
where
$\eta\equiv (4\pi M^{-n+1}_P/\epsilon )^{-1/(n-3)}$.

First, let us examine the case where the horizon is the particle horizon, i.e., $r=R_p$.
From Eq. (\ref{Rp}) and the above equation one can obtain
\beq
\int^R_0 \frac{dR'}{HR'^2}=\frac{\eta H^{\frac{2}{n-3}}}{R}.
\eeq
Differentiating the above equation with $R$ we obtain
\beq
\label{diff}
\frac{1}{HR^2}=\eta \frac{d}{dR} \left(\frac{H^{\frac{2}{n-3}}} {R} \right).
\eeq
Setting $H^{-1}\equiv \alpha R^x$
we rewrite the above equation as
\beqa
\alpha R^{x-2}&=&\eta  \frac{d}{dR} \left( (\alpha R^x)^{\frac{-2}{n-3}} {R} \right)\no
&=&\eta \alpha^{\frac{-2}{n-3}}\left(\frac{-2x}{n-3}-1\right) R^{\frac{-2x}{n-3}-2}.
\eeqa
Comparing both sides, one can find that
$n=1$. Thus, $\eta$ turns out to be dimensionless in this case. One can also note that $x=1+1/\eta$.
Thus $t^{-2}\propto H^2\propto R^{-2x}\propto R^{-2 (1+\frac{1}{\eta})}$
and, hence, $R\propto t^{\frac{\eta}{\eta+1}}$. Therefore,
the particle horizon does not give rise to an accelerating expansion and
there is no dark energy in this case.

Now we move on the case with the future event horizon.
 Repeating the same procedure using $R_h$ for $r$ instead of $R_p$
(i.e., $r=R_h$) one can obtain $n=1$, $x=1-1/\eta$
and $R\propto t^{\frac{\eta}{\eta-1}}$ which shows
accelerating expansion as in HDE model.
This difference came from the position of $R$ in the integral limit  of  Eq. (\ref{Rp}) and  Eq. (\ref{Rh}).
Therefore, we confirm that the future event horizon is a natural
horizon for dark energy in our model.
Since $n=1$, our dark energy has the mathematical form of HDE,
i.e., $\rho\propto R^{-2}_h$ in general.
Comparing Eq. (\ref{rho}) with the holographic dark energy;
\beq
\rho_\Lambda=\frac{3\epsilon M^2_P}{4 \pi R_h^2}=\frac{3 d^2 M_P^2  }{ R_h^2 },
\eeq
one can find that
\beq
d=\sqrt{\frac{\epsilon}{4\pi }}={\eta}.
\eeq
According to observations~\cite{zhang-2007} $d\simeq 1$, thus
$\epsilon\simeq 4\pi$.

How can we verify  our model by current observations?
Once we obtain $\rho_\Lambda$,
the negative pressure $p_\Lambda$ can be  derived from the
cosmological energy-momentum conservation equation
as usually done in holographic dark energy models~\cite{1475-7516-2004-08-013}.
From the Friedmann equation with perfect fluid
having  stress-energy tensor
\beq
\label{Tperfect}
T_{\mu\nu}=(\rho_\Lambda+p_\Lambda) U_\mu U_\nu- p_\Lambda g_{\mu\nu},
\eeq
where $U^{\mu}U_{\mu}=1$, one can derive the cosmological energy-momentum conservation equation;
\beq
\label{p}
p_\Lambda=\frac{d(R^3\rho_\Lambda)}{-3 R^2 dR }.
\eeq
From this, one can notice that
 a perfect fluid with total energy increasing as
the universe expands has  effective negative pressure.
From Eq. (\ref{p}) one can find that the present equation of state for HDE in Eq. (\ref{holodark2})
 is  ~\cite{li-2004-603,1475-7516-2004-08-006}
\beq
\label{omega}
w^0_\Lambda =-\frac{1}{3} \left(1+\frac{2\sqrt{\Omega^0_\Lambda} }{d}\right)=
-\frac{1}{3} \left(1+2\sqrt{\frac{4 \pi   \Omega^0_\Lambda}{\epsilon} }\right),
\eeq
and
the change of the equation of state is given by ~\cite{1475-7516-2004-08-006}
\beqa
\label{omega3}
\frac{d w_{\Lambda}(z)}{dz} &=&
\frac{\sqrt{\Omega_\Lambda} \left( 1 - \Omega_\Lambda  \right)}{3d(1+z)} \left(1+\frac{2\sqrt{\Omega_\Lambda} }{d}\right) \no
&=& \frac{\sqrt{4\pi\Omega_\Lambda} \left( 1 - \Omega_\Lambda  \right)}{3\sqrt{\epsilon}(1+z)}\left(1+\frac{2\sqrt{4 \pi\Omega_\Lambda} }{\sqrt{\epsilon}}\right)
\eeqa
where  $z\equiv 1/R-1$ is the red shift parameter  and $\Omega^0_\Lambda\simeq 0.73$ is the observed present  value of
the density parameter
$\Omega_\Lambda$ of the dark
energy. (Here we set the current scale factor $R_0=1$.)
To compare the results with observations it is useful to represent the above equation as a function of $1-R$
using the relation;
\beqa
\label{dw}
\frac{dw_\Lambda}{dz}&=&\frac{d(1-R)}{dz}\frac{dw_\Lambda}{d(1-R)}\no
&=&R^2\frac{dw_\Lambda}{d(1-R)}= \frac{1}{(1+z)^2}
\frac{dw_\Lambda}{d(1-R)}.
\eeqa
Then, for $z\ll 1$, $
w_\Lambda (z) \simeq w^0_\Lambda+w_1 (1-R)$,
where
\beq
\label{w1}
 w_1\equiv \frac{d w_{\Lambda}}{d(1-R)}|_{R=1}= \frac{d w_{\Lambda}(z)}{dz}|_{z=0}.
 \eeq
Now, we can compare predictions of our theory with recent observations~\cite{xia-2007,Zhao:2006qg}.
Inserting the observed $w^0_\Lambda = -1.03 \pm 0.15$ into Eq. (\ref{omega}), one can obtain $\epsilon=8.4^{+5.2}_{-2.7}$.
From this we also obtain $w_1=0.29^{+0.11}_{-0.10}$ using Eq. (\ref{omega3})  and Eq. (\ref{dw}), which are comparable to the
observed value, although the observational uncertainty is still large.
The near future observations will give stronger constraints on
 $w^0_\Lambda$ and $w_1$, and hence, on  $\epsilon$.
(Note that the specific  model considered in section II and III already  predicts $d\simeq1$ and hence  $\epsilon\simeq 4\pi$.)

\section{Discussion}

The cosmic coincidence problem  can  be also  easily solved in our model using the result of
\cite{mycoin,li-2004-603}, if there was an ordinary inflation with the number of e-folds $N\simeq 65$ at the very
early universe driven by some inflaton fields,
because our model gives dark energy in the form of  HDE.

Where does the energy erasing information come from?
 A similar issue exists in the inflation theory
regarding the expansive reproduction of the  vacuum energy during the inflation.
Since the information is erased as the horizon expands, for an inside observer,
the horizon plays a role of the cylinder in Fig.1.
And this horizon expands due to the cosmic expansion after the inflation.
Thus, one can imagine that the energy comes from `kinetic' energy of the universe originated
from the inflationary expansion.
Alternatively, one can also imagine the energy is borrowed from  negative gravitational
energy as in the idea of the zero-energy universe~\cite{TRYON1973}.
In the both scenarios dark energy and inflation have a deep connection.
However, note that the energy alone is
usually not conserved in general relativity.
What we can safely rely on is the energy momentum conservation in Eq. (\ref{p}).

It is still possible that at the earlier universe the dark entropy and the temperature are not  simple
power-law functions of $R_h$. Furthermore, another horizon besides the event horizon could
give an accelerating universe when we include the interaction between dark energy and ordinary matter.
 In these cases
our theory may give a dark energy model deviated from the simple holographic dark energy model
and   be free from the known difficulties of HDE model.

It is also interesting that the cosmic Hawking radiation or the energy from Landauer's principle
 may be simulated on the acoustic horizons ~\cite{PhysRevLett.46.1351,PhysRevD.58.064021} or optical black holes~\cite{opticalBH} in the future.

Our theory require neither exotic matter, fine-tuning of
potential nor modification of gravity.
Assuming  the holographic principle, Landauer's principle and
Gibbons-Hawking temperature, one can well describe the observed dark energy.
Our work indicates that the solution of the dark energy problem
may not rely on exotic materials or radical new physics but
the holographic principle and
new aspects of familiar quantum physics such as entanglement and  Landauer's principle.

\section*{acknowledgments} This work was partly supported by the IT R\&D program of MIC/IITA
[2005-Y-001-04 , Development of next generation security technology]
 (J.-W.L.) and the Korea Research Foundation Grant funded by Korea Government
 (MOEHRD, Basic Research Promotion Fund)(KRF-2006-312-C00095;J.J. L).


\end{document}